\documentclass[preprint,12pt]{elsarticle}
\usepackage{newtxtext,newtxmath}
\biboptions{numbers,sort&compress}
\usepackage{sections/mypreamble}
\usepackage{times}
\usepackage{amsmath}
\usepackage{epstopdf}
\usepackage{setspace}
\usepackage{float}
\usepackage{xcolor}

\usepackage{hyperref}                       % For clickable links in PDF

\hypersetup{
    colorlinks=true,                        % Enable colored links instead of boxes
    linkcolor=blue,                         % Color for internal links (sections, etc.)
    citecolor=green,                        % ✅ Green color for citation numbers
    filecolor=magenta,                      % Color for file links (if used)
    urlcolor=cyan,                          % Color for external URLs
    pdftitle={Your Paper Title},            % Sets the PDF metadata title
    pdfpagemode=FullScreen                  % Opens the PDF in full-screen mode
}

\begin{document}
\begin{frontmatter}

    \title{Online Censoring-Based Widely Linear Total Least lncosh Method for Improved Power System Frequency Estimation}

    \author[label1]{Haiquan Zhao \corref{mycorrespondingauthor}}
    \ead{ hqzhao@swjtu.edu.cn}
    \author[label1]{Kaleab Derbew Abebe}
    \author[label1]{Yi Peng}
    
    \affiliation[label1]{organization={Haiquan Zhao, Kaleab Derbew Abebe and Yi Peng are with the Key Laboratory of Magnetic Suspension Technology and Maglev Vehicle, Ministry of Education},
             addressline={Southwest Jiaotong University},
             city={Chengdu 610031},
             country={China}}

    \cortext[mycorrespondingauthor]{Corresponding author}
\doublespacing
    \begin{abstract}
        Recently, under the presumption of a noise-free input, the augmented complex least lncosh (ACLlncosh) method was introduced for a power system frequency estimate and showed robust performance when impulsive noise polluted the output signal. However, in practical terms, noise often contaminates input signals, which drastically reduces the efficiency of the ACLlncosh method. To enhance robustness against noisy input-output while maintaining resilience to impulsive noise in the output signal, this paper proposed an online censoring-based widely linear total least lncosh (OC-WL-TLlnC) method. This method improves performance under both balanced and unbalanced settings by filtering out less valuable data via online censoring, hence reducing the computing burden. Furthermore, a variable parameter approach is incorporated to accelerate convergence and improve steady-state accuracy, thereby ensuring adaptability to dynamic power system conditions. The proposed methods significantly enhance frequency estimate performance by addressing the constraints of current techniques and offering a computationally efficient, noise-resilient solution for real-time power system monitoring.
    \end{abstract}

\end{frontmatter}

    \doublespacing

    %% The lineno packages adds line numbers. Start line numbering with
    %% \begin{linenumbers}, end it with \end{linenumbers}. Or switch it on
    %% for the whole article with \linenumbers.
    %% \usepackage{lineno}

    \acresetall
    \section{Introduction}\label{sec:introduction}
Frequency is one of the most important parameters for assessing the operational state of a power grid. Consequently, accurate and rapid frequency estimation is a key function of wide-area measurement systems, as it reflects the balance between generation and load in real time. Over the years, various methods have been proposed in both signal processing and power system domains \cite{1,2,3,4,5}. While most existing approaches \cite{6,7,8,9,10,11} are designed for single-phase systems and perform well under balanced conditions, they often struggle to capture the asymmetries and inter-phase interactions in three-phase power systems under unbalanced conditions. 

To address this, recent research \cite{12,13,14}, notably the work by Xia and Mandic in 2012 \cite{13}, has introduced widely linear modeling techniques for frequency estimation under unbalanced conditions. These methods use Clark's $\alpha \beta$ transformation to combine three-phase voltages into a single complex-valued signal, making it easier to analyze the system using advanced tools from complex signal processing, especially those based on augmented second-order statistics. 

Among the different techniques developed using this framework, adaptive algorithms are widely favored due to their simplicity, low computational cost, and resilience to noise and harmonics \cite{15}. 
However, many of these methods are based on minimizing the mean square error (MSE), which considers only second-order error statistics. As a result, their performance tends to decline in non-Gaussian environments, where more robust error criteria are often required. 

In real-world power systems, impulsive noise often appears during power transmission or while collecting data \cite{14}. This often happens with things like power line communication systems \cite{16} or electric arc furnaces \cite{17}. But since least mean square (LMS) algorithms rely on the MSE criterion, which assumes the noise is Gaussian, they do not perform well when the noise has a non-Gaussian distribution \cite{18}. To overcome this issue, several algorithms have been proposed \cite{19,20}. However, the performance of LMS-based algorithms deteriorates under the errors-in-variables (EIV) model, in which the input signal is affected by noise as well \cite{21}. 

Given these challenges, several researchers have explored the use of the complex total least squares (CTLS) method to address issues that arise when noise also affects the input signals, particularly in systems with Gaussian noise \cite{22}. On the other hand, TLS requires more computation, and its performance degrades when the noise is non-Gaussian \cite{23,24}. The augmented complex least lncosh (ACLlncosh) algorithm mitigates impulsive output noise using a lncosh cost function that does not rely on a specific noise distribution \cite{18}. Still, it assumes that the input is noise-free. This assumption can significantly degrade performance in EIV scenarios. Accuracy often drops, and steady-state errors may occur, particularly under dynamic or unbalanced conditions, limiting its practical use in real-time applications.

To address both computational and robustness concerns, recent studies have proposed strategies like online censoring (OC), which discards statistically uninformative data in real time \cite{25}. Building on this idea, the OC strategy evaluates each incoming sample as it arrives, updating the parameters only when the data is informative. This reduces computational cost without sacrificing performance, making it especially useful in engineering applications \cite{26}. This method has been shown to significantly decrease storage requirements and computational complexity in several real-time applications \cite{27,28}. Despite its promising features, OC has not yet been explored for power system frequency estimation. 

\setlength{\parskip}{0pt}

Based on these observations, there is a strong need for frequency estimation algorithms that can resist impulsive noise and input-output measurement errors while running fast enough for real-time applications with large streaming data. Current methods either assume noise free-inputs, perform poorly under non-Gaussian or unbalanced conditions, or require high computational effort, which limits their use in modern power systems. Addressing these limitations is essential for accurate monitoring, fault detection, and maintaining system stability under dynamic conditions. 

This paper introduces the OC-WL-TLlnC algorithm, a robust complex-domain TLS method with online censoring, designed to overcome the limitations of the Llncosh approach. The algorithm focuses on the most informative data and limits the effect of noise and outliers. Using widely linear modeling, it can handle non-circular signals that commonly appear in three-phase unbalanced systems. Theoretical analysis confirms stability and convergence, and simulations show improved accuracy and robustness compared to traditional methods, making it well suited for real-time power system applications.

\textbf{Notations}: For the three-phase frequency estimation task, several mathematical models are presented. In the beginning, some standard definitions of mathematical notation are provided. $(\cdot)^T$ is transposed, $(\cdot)^H$ is conjugate transposition, $j^2 = -1$, $(\cdot)^*$ is conjugate, $|\cdot|$ is an absolute value operation, $\operatorname{E}(\cdot)$ denotes the expectation value, and $\|\cdot\|$ represents the 2-norm operator.

    \section{Problem statement}
\label{sec2}
\subsection{Three-phase system modeling for frequency estimation}
\label{subsec1}
The voltage signals in a three-phase system can be represented through a discrete-time model given below, as described in \cite{12}
\begin{align} 
	v_{a}(k) &= V_a(k)\cos\left(\omega k\Delta T + \theta\right) \nonumber \\
	v_b(k) &= V_b(k)\cos\left(\omega k\Delta T + \theta + \Delta\theta_b - \frac{2\pi}{3}\right) \nonumber \\
	v_c(k) &= V_c(k)\cos\left(\omega k\Delta T + \theta + \Delta\theta_c + \frac{2\pi}{3}\right) \label{eq:1}
\end{align}
where $V_a(k)$, $V_b(k)$, and $V_c(k)$ represent the amplitudes of the phase voltages at certain moments in time $k$, $f_s = 1/\Delta T$ indicates the sampling frequency, $\Delta T$ denotes the sampling period, $\theta$ indicates the initial phase of the fundamental component, and $\omega = 2\pi f$ denotes angular frequency in which the fundamental system frequency $f$ is embedded. A balanced system state is represented by $\Delta\theta_b$ and $\Delta\theta_c$, which are the phase angle deviations of the phases $b$ and $c$, respectively. Based on this, $v_a(k)$, $v_b(k)$, and $v_c(k)$ are then transformed into the $\alpha\beta$ coordinate frame using Clarke's transformation, as shown below 
\begin{align}\label{eq:2} \begin{bmatrix} v_0 \\ v_\alpha \\ v_\beta \end{bmatrix}=
	\sqrt\frac{2}{3}   
	\begin{bmatrix} \frac{\sqrt{2}}{2} & \frac{\sqrt{2}}{2} & \frac{\sqrt{2}}{2} \\ 1 & -\frac{1}{2} &  -\frac{1}{2} \\ 0 & \frac{\sqrt{3}}{2} & -\frac{\sqrt{3}}{2} \\ \end{bmatrix}  
	\begin{bmatrix} v_a(k)\\ v_b(k)\\  v_c(k) \end{bmatrix}\end{align}
    
The complex voltage, $v(k)$, which represents the three-phase system, is derived from \eqref{eq:2} as follows
\begin{align}\label{eq:3}
	v(k) &= v_\alpha(k)+jv_\beta(k) 
\end{align}

From the relationships given in equations \eqref{eq:2} and \eqref{eq:3}, the system voltage in its complex form, $v(k)$, is derived from equation \eqref{eq:3} as follows
\begin{equation}\label{eq:4}
v(k) = A(k) e^{j(\omega k \Delta T + \theta)} + B(k) e^{-j(\omega k \Delta T + \theta)} \tag{4}
\end{equation}
\noindent where
\begin{align}\label{eq:5}\tag{5}
A(k) &= \frac{\sqrt6}{6}\left(V_a(k)+V_b(k)e^{j \Delta \theta_b}+V_c(k)e^{j \Delta \theta_c}\right) 
\end{align}
and
\begin{equation}\label{eq:6}\tag{6}
\hspace*{-\parindent}
\scalebox{0.96}{$
B(k) = \frac{\sqrt{6}}{6} \left( V_a(k) + V_b(k)e^{-j\left(\Delta \theta_b + \frac{2\pi}{3}\right)} 
+ V_c(k)e^{-j\left(\Delta \theta_c - \frac{2\pi}{3}\right)} \right)
$}
\end{equation}

It is observed in a balanced three-phase power system that the voltages across each phase are identical, with $V_a(k)$, $V_b(k)$, and $V_c(k)$ being equal, and the phase angles $\theta_b$ and $\theta_c$ are both zero. Consequently, certain conclusions can be drawn \eqref{eq:4} becomes
\begin{align}\label{eq:7}
	v(k) &= \sqrt{\frac{3} {2}} V_a(k)e^{j(\omega k \Delta T + \theta)}\tag{7}
\end{align}
\noindent and
\begin{align}
\label{8} A(k) &= \sqrt{\frac{3}{2}}V_a(k) \tag{8} \\
\label{9} B(k) &= 0 \tag{9}
\end{align}

\subsection{EIV model}
\label{subsec1a}
The problem is first framed using a linear system as follows
\begin{equation} \label{eq:10}\tag{10}
\tilde{y}(k) = \boldsymbol{w}_o\boldsymbol{x}(k)
\end{equation}
\noindent where $ \boldsymbol{w}_o \in \mathbb{C}^{L \times 1}$ is the weight vector to be assessed by the system. $\tilde{y}(k) \in \mathbb{C}$ represents the output signal, whereas $\boldsymbol{x}(k) \in \mathbb{C}^{L \times 1}$ denotes the input signal. In a real-world scenario, both signals exhibit noise. Thus, the behavior of the system can be characterized as follows in terms of the EIV model:
\begin{align} \label{eq:11}\tag{11}
	\tilde{d}(k) &= \tilde{y}(k) + p(k) \\
	\boldsymbol{\tilde{x}}(k)&= \boldsymbol{x}(k) + q(k) \label{eq:12}\tag{12}
\end{align}
where $p(k) \in \mathbb{C}$ is the output noise, and $\boldsymbol{q}(k) \in \mathbb{C}^{L \times 1}$ is the input noise, both modeled as complex Gaussian white noise. The variances of the real and imaginary parts are $\sigma_{q}^2$ and $\sigma_p^2$, respectively.  Accordingly, the error signal is given by
\begin{equation} \label{eq:13}\tag{13}
	e(k) = \tilde{d}(k) - y(k) = \tilde{d}(k) - \boldsymbol{w}(k)\boldsymbol{\tilde{x}}(k)
\end{equation}
\noindent where $y(k)$ denotes the output signal generated by the adaptive FIR filter, and $\boldsymbol{w}(k)$ corresponds to its associated weight vector.
\section{Frequency estimation algorithms}
\label{sec3}
\subsection{Complex signal-based TLlnC algorithm}\label{subsec1b}
A detailed derivation and analysis of the conventional complex least lncosh (CLlncosh) algorithm for power system frequency estimation can be found in \cite{18}. Based on this, the output $ \hat{v}(k+1) $, the estimation error $ e(k) $, and the update rules for the filter coefficient $ \boldsymbol{w}(k) $ in the strictly linear CLlncosh (SL-CLlncosh) method are given below
\begin{align}\label{eq:14}
	\hat{v}(k+1) &= \boldsymbol{v}(k)\boldsymbol{w}(k)\tag{14}
\end{align}
\begin{equation}\label{eq:15}
    e(k) = v(k+1) - \hat{v}(k+1)\tag{15}
\end{equation}
\begin{equation} \label{eq:16} \tag{16}
   \boldsymbol{w}(k+1) = \boldsymbol{w}(k) -  \mu \left( \nabla_{\boldsymbol{\boldsymbol{w}}^*} J_{Incosh} \right)
\end{equation}

\noindent where $\boldsymbol{\boldsymbol{v}}(k)$ indicates the complex value input voltage signal, $\boldsymbol{v}(k+1)$ represents the desired voltage signal of the system, $\mu$ denotes the step-size, and $\nabla_{\boldsymbol{\boldsymbol{w}}^*}$ denotes the complex gradient of $J_{\text{Incosh}}$ about $\boldsymbol{\boldsymbol{w}}$. Considering that consecutive complex voltage samples $\boldsymbol{v}(k)$ and $\boldsymbol{v}(k+1)$ serve as the input and desired signal, respectively, the final weight update expression in \eqref{eq:16} is derived as follows
\begin{equation}
\boldsymbol{w}(k+1) = \boldsymbol{w}(k) + \mu \tanh(\lambda(k) e(k)) \boldsymbol{v}^*(k) \tag{17}
\end{equation}

Under balanced conditions, the three-phase power signal exhibits circularity, allowing the SL-CLIncosh method in \eqref{eq:14} to estimate the system frequency effectively based on the model in \eqref{eq:4}. Thus, the target voltage signal $\boldsymbol{v}(k+1)$ is iteratively derived from \eqref{eq:4} as follows
\begin{equation}\label{eq:18}\tag{18}
\begin{aligned}
v(k+1) &= A(k+1)e^{j(\omega(k+1)\Delta T + \theta)} \\
&= \left[A(k)e^{j(\omega k \Delta T + \theta)}\right]e^{j\omega \Delta T} = \boldsymbol{v}(k)e^{j\omega \Delta T}
\end{aligned}
\end{equation}

By comparing \eqref{eq:18} to the SL-CLlncosh approach in \eqref{eq:14}, it becomes evident that the filter weight $\boldsymbol{w}(k)$ closely follows the form \( e^{j\omega \Delta T} \). This relationship makes it possible to extract the system's instantaneous frequency. Using the estimated voltage $\hat{v}(k+1)$, the fundamental frequency can then be determined as
\begin{equation}
\hat{f} = \frac{1}{2\pi \Delta T} \sin^{-1} \left( \Im(\boldsymbol{w}(k) \right)) \tag{19}
\end{equation}

Due to unequal amplitudes in $\boldsymbol{v}_a(k)$, $\boldsymbol{v}_b(k)$, and $\boldsymbol{v}_c(k)$ along with a nonzero $B(k)$, the linear model in \eqref{eq:14} becomes invalid. As discussed in \cite{18}, the conventional linear CLlncosh model cannot effectively handle unbalanced conditions. In such cases, an SL structure causes the weight coefficient to oscillate at twice the system frequency in a steady state, which in turn distorts the estimated frequency $\hat{f}(k)$, as shown below
\begin{equation}
\boldsymbol{w}(k) = \boldsymbol{w}\left(k + \frac{1}{2f\Delta T}\right) \tag{20}
\end{equation}

To overcome this issue, a more suitable solution for an unbalanced three-phase power system is the ACLIncosh algorithm, which is based on the WL model and can be expressed as
\begin{align}
\hat{\boldsymbol{v}}(k+1) &= \boldsymbol{v}(k) \boldsymbol{h}(k) + \boldsymbol{v}(k)^*\boldsymbol{g}(k) \label{eq:21}\tag{21} \\
e(k) &= \boldsymbol{v}(k+1) - \hat{\boldsymbol{v}}(k+1) \label{eq:22}\tag{22} \\
\boldsymbol{h}(k+1) &= \boldsymbol{h}(k) - \mu \left( \nabla_{\boldsymbol{h}^*} J_{\text{Incosh}} \right) \label{eq:23}\tag{23} \\
\boldsymbol{g}(k+1) &= \boldsymbol{g}(k) - \mu \left( \nabla_{\boldsymbol{g}^*} J_{\text{Incosh}} \right) \label{eq:24}\tag{24}
\end{align}

From the above, $\boldsymbol{h}(k)$ and $\boldsymbol{g}(k)$ signify the standard and conjugate components of the weight vector at time $k$. The gradients $\nabla_{\boldsymbol{h}^*}$ and $\nabla_{\boldsymbol{g}^*}$ are taken with respect to the standard and conjugate weights in the Llncosh cost function of $J_{\text{Incosh}}$.

Accordingly, the weight update in \eqref{eq:23} and \eqref{eq:24} are derived as
\begin{align}
\boldsymbol{h}(k+1) &= \boldsymbol{h}(k) + \mu \tanh(\lambda(k) e(k) )\boldsymbol{v}^*(k) \label{eq:25}\tag{25} \\
\boldsymbol{g}(k+1) &= \boldsymbol{g}(k) + \mu \tanh(\lambda(k) e(k)) \boldsymbol{v}(k)\label{eq:26}\tag{26}
\end{align}

Using the ACLIncosh model in \eqref{eq:21} along with the updates in \eqref{eq:25} and \eqref{eq:26}, the fundamental frequency can be estimated as
\begin{equation}
\hat{f} = \frac{1}{2\pi \Delta T} \tan^{-1} \left\{ \frac{\Im \left[ \boldsymbol{h}(k) + a^1(k) \boldsymbol{g}(k) \right]}{\Re \left[ \boldsymbol{h}(k) + a^1(k) \boldsymbol{g}(k) \right]} \right\} \tag{27}
\end{equation}

In this context, the discriminant $a^1(k)$ can be expressed as
\begin{equation}
a^1(k) = \frac{-j \Im(\boldsymbol{h}(k)) + j \sqrt{\Im^2(\boldsymbol{h}(k)) - |\boldsymbol{g}(k)|^2}}{\boldsymbol{g}(k)} \tag{28}
\end{equation}

\subsection{Frequency estimation utilizing the proposed online censoring-based CTLInC and WL-TLInC methods}\label{subsec1c}
Building on prior real-valued formulations, the proposed technique integrates the complex domain Llncosh (CLlncosh) model and an OC criterion into the TLS framework. The corresponding cost function is derived as follows
\begin{equation} \label{eq:29} \tag{29}
J_{CTLlnC} = E \left[ \frac{1}{\lambda} \ln \left( \cosh \left( \lambda 
|e^o(k)| \right) \right) \right]
\end{equation}
\noindent here: $\lambda $ is a non-negative parameter, $ \ln(\cdot) $ denotes the natural logarithm, 
$
\cosh(\lambda |e^o(k)|) = \frac{\exp(\lambda |e^o(k)|) + \exp(-\lambda |e^o(k)|)}{2},
$
and the overall error is expressed as
\begin{align}\label{eq:30}
	e^o(k) = e(k)/\|\boldsymbol{\bar{w}}\|
\tag{30}
\end{align}
here, $\boldsymbol{\bar{w}}=[w(k), \sqrt{\gamma }]^T$, ${\gamma }={\sigma}_{p}^{2}/{\sigma}_{q}^{2}$ denotes the ratio between the output and input noise variances.

Considering the online censoring rule, the truncated cost function $ J^{(\tau)}_{CTLlnC} $ is first defined based on \cite{26} and then constructed by incorporating the $ J_{CTLlnC} $ model formulation, as detailed below
\begin{equation}\label{eq:31}
\begin{aligned}
     J^{(\tau)}_{CTLlnC} := 
    \begin{cases}
        E \left[ \frac{1}{\lambda} \ln \left( \cosh(\lambda |e^o(k)| \right) \right], & \text{if } \tau \sigma_p \leq |e^o(k)| < \tau^o \sigma_p \\
        0, & \text{if } |e^o(k)| < \tau \sigma_p
    \end{cases}
\end{aligned}
\tag{31}
\end{equation}

\noindent where $\tau > 0$, $\sigma_p$, $\tau^o > \tau$ denotes the censoring threshold, the standard deviation of $p(k)$, and the additional threshold. Using the stochastic gradient descent technique, the OC strategy may be minimized as
\begin{equation} \label{eq:32} 
\begin{aligned}
     \hat{\nabla}_{\boldsymbol{w}^*} J^{(\tau)}_{CTLlnC} = 
     \left\{ 
    \begin{array}{ll}
        \hat{\nabla}_{\boldsymbol{w}^*} J_{CTLlnC}, & \text{if } \tau \sigma_p \leq |e^o(k)| < \tau^o \sigma_p \\
        0, & \text{if } |e^o(k)| < \tau \sigma_p 
    \end{array}
    \right.
\end{aligned}
\tag{32}
\end{equation}
\noindent where
\begin{equation} \label{eq:33}
\begin{split}
    \hat{\nabla}_{\boldsymbol{w}^*} J_{CTLlnC} = -\tanh \left( \lambda |e^o(k)| \right) \cdot \frac{e^o(k)}{|e^o(k)|} \cdot 
    \left[ \frac{\boldsymbol{v}^*(k)}{\|\boldsymbol{\bar{w}}\|} + \frac{\boldsymbol{w}(k)e^*(k)}{\|\boldsymbol{\bar{w}}\|^3} \right] 
\end{split}\tag{33}
\end{equation}
\noindent where $\tanh \left( \lambda |e^o(k)| \right)$ represents the hyperbolic tangent function, defined as
\begin{equation}\label{eq:34}
\tanh(\lambda |e^o(k)|) = \frac{e^{(\lambda |e^o(k)|)} - e^{(-\lambda |e^o(k)|)}}{e^{(\lambda |e^o(k)|)} + e^{{(-\lambda |e^o(k)|)}}} \tag{34}
\end{equation}

The weight update rule for the proposed OC-CTLlnC algorithm is derived based on the gradient vector defined in \eqref{eq:33} as follows
\begin{equation} \label{eq:w_update}
\begin{aligned}
    \boldsymbol{w}(k+1): =  \left\{
    \begin{array}{ll}
        \boldsymbol{w}(k) + \mu \hat{\nabla}_{\boldsymbol{w}^*} J_{CTLlnC}, & \text{if } \tau \sigma_p \leq |e^o(k)| < \tau^o \sigma_p \\
        \boldsymbol{w}(k), & \text{if } |e^o(k)| < \tau \sigma_p
    \end{array}
    \right.
\end{aligned} \tag{35}
\end{equation}
\noindent where $\mu$ signifies the step size, and $\hat{\nabla}\boldsymbol{w}^{*}J_{CTLlnC}$ represents the gradient of the cost function.
Following the principles of the adaptive variable parameter method in \cite{18}, the time-varying parameter $\lambda(k)$ is adjusted as follows
\begin{equation} \label{eq:36} \tag{36}
\lambda(k+1) = \beta \lambda(k) + (1-\beta)|e^o(k)|^2
\end{equation}
\noindent here, $\beta \in (0, 1)$ denotes the forgetting factor. Additionally, adapting $\lambda(k)$ based on the prediction error enhances the tracking capability of the method.
\subsection{Proposed OC-WL-TLlnC method}\label{subsec3}
In contrast, to reduce or eliminate the frequency estimation fluctuation observed in conventional linear estimators and achieve accurate results under unbalanced conditions, a WL estimation approach is introduced. The OC-WL-TLlnC function, used for estimating the frequency of a three-phase power system, is formulated as follows
\begin{equation} \label{eq:37}
	\hat{{v}}(k+1) = \boldsymbol{v}(k)\boldsymbol{h}(k) + \boldsymbol{v}^*(k)\boldsymbol{g}(k)\tag{37}
\end{equation}
\begin{equation} \label{eq:38}
	\hat{e}(k) = v(k+1) - \hat{v}(k+1)\tag{38}
\end{equation}
\begin{equation} \label{eq:39}
	\boldsymbol{h}(k+1) = \boldsymbol{h}(k) + \mu \hat{\nabla}_{\boldsymbol{\boldsymbol{h}}^*} \hat{J}_{WL-TLlnC}^{}(|\hat{e}^o(k)|)\tag{39}
\end{equation}
\begin{equation} \label{eq:40}
	\boldsymbol{g}(k+1) = \boldsymbol{g}(k) + \mu \hat{\nabla}_{\boldsymbol{\boldsymbol{g}}^*} \hat{J}_{WL-TLlnC}^{}(|\hat{e}^o(k)|)\tag{40}
\end{equation}

The overall error in the WL model is articulated by including \eqref{eq:38} into \eqref{eq:30}, where $\boldsymbol{h}(k)$ denotes the standard weight and $\boldsymbol{g}(k)$ the conjugate weight in the WL filtering, respectively, as
\begin{equation} \label{eq:41}
	\hat{e}^o(k) = \frac{{v}(k+1) - \boldsymbol{v}(k)\boldsymbol{h}(k) - \boldsymbol{v}^*(k)\boldsymbol{g}(k)}{\sqrt{|\boldsymbol{h}|^2 + |\boldsymbol{g}|^2+ \gamma} }\tag{41}
\end{equation}

Based on \eqref{eq:29}, the WL-Llncosh criterion is integrated into the TLS framework, leading to a new cost function that represents the overall error, formulated as follows
\begin{equation} \label{eq:42}
\hat{J}_{WL-TLlnC} = E \left[ \frac{1}{\lambda} \ln \left( \cosh \left( \lambda |\hat{e}^o(k)| \right) \right) \right] \tag{42}
\end{equation}

Consequently, incorporating \eqref{eq:42}, the resilient truncated cost function $\hat{J}_{WL-TLlnC}^{\tau}$ articulated as
\begin{equation}\label{eq:43}
\begin{aligned}
    \hat{J}^{(\tau)}_{WL-TLlnC} := 
    \begin{cases}
        E \left[ \frac{1}{\lambda} \ln \left( \cosh \left( \lambda |\hat{e}^o(k)| \right) \right) \right], & \text{if } \tau \sigma_p \leq |\hat{e}^o(k)| < \tau^o \sigma_p \\
        0, & \text{if } |\hat{e}^o(k)| < \tau \sigma_p
    \end{cases}
\end{aligned}
\tag{43}
\end{equation}

By differentiating \eqref{eq:43} with respect to $\boldsymbol{h}^*$ and $\boldsymbol{g}^*$, the following update rules are obtained as follows
\begin{equation}\label{eq:44}
\begin{aligned}
   &\nabla_{\boldsymbol{h}^*} \hat{J}_{WL-TLlnC} (|\hat{e}^o(k))| = \frac{\partial \hat{J}_{WL-TLlnC}}{\partial \boldsymbol{h}^*} = -\tanh(\lambda |\hat{e}^o(k)|) \cdot \\  
   & \frac{\hat{e}^o(k)}{|\hat{e}^o(k)|} \cdot
   \left[ \frac{\boldsymbol{v}^*(k)}{\sqrt{|\boldsymbol{h}|^2 + |\boldsymbol{g}|^2 + \gamma}} + \frac{\boldsymbol{h}(k)\hat{e}^{o^*}(k)}{|\boldsymbol{h}|^2 + |\boldsymbol{g}|^2 + \gamma} \right]
\end{aligned} \tag{44}
\end{equation}
\begin{equation}\label{eq:45}
\begin{aligned}
   &\nabla_{\boldsymbol{g^*}} \hat{J}_{WL-TLlnC} (|\hat{e}^o(k)|)=  \frac{\partial \hat{J}_{WL-TLlnC}}{\partial \boldsymbol{g}^*} = -\tanh(\lambda |\hat{e}^o(k)|)\cdot \\  
   & \frac{\hat{e}^o(k)}{|\hat{e}^{o}(k)|}  \cdot
   \left[ \frac{\boldsymbol{v}(k)}{\sqrt{|\boldsymbol{h}|^2 + |\boldsymbol{g}|^2 + \gamma}} + \frac{\boldsymbol{g}(k)\hat{e}^{o^*}(k)}{|\boldsymbol{h}|^2 + |\boldsymbol{g}|^2 + \gamma} \right]
\end{aligned} \tag{45}
\end{equation}

Similarly, in the prior section, the reduction of \eqref{eq:43} yields the updated equation for the weight of the OC-WL-TLlnC, as follows
\begin{equation} \label{eq:46}
\begin{aligned}
   &\text{ if } \tau \sigma_p \le |\hat{e}^o(k)| < \tau^o \sigma_p, \quad 
\begin{cases}
        \boldsymbol{h}(k+1) := \boldsymbol{h}(k) + \mu \hat{\nabla}_{\boldsymbol{h}^*} \hat{J}_{WL-TLlnC}(|\hat{e}^o(k)|) \\
        \boldsymbol{g}(k+1) := \boldsymbol{g}(k) + \mu\hat{\nabla}_{\boldsymbol{g}^*} \hat{J}_{WL-TLlnC}(|\hat{e}^o(k)|)
    \end{cases} \\
    &\text{ if } |\hat{e}^o(k)| < \tau \sigma_p, \quad \begin{cases}
        \boldsymbol{h}(k+1) := \boldsymbol{h}(k) \\
       \boldsymbol{g}(k+1) := \boldsymbol{g}(k)
    \end{cases}
\end{aligned}\tag{46}
\end{equation}
\noindent where the step size $\mu$, analogous to that used in the OC-CTLlnC method, is applied to the complex gradients $\mu \nabla_{\boldsymbol{h}^*} \hat{J}_{WL-TLlnC} (|\hat{e}^o(k)|)$ and $\mu \nabla_{\boldsymbol{g}^*} \hat{J}_{WL-TLlnC} (|\hat{e}^o(k)|)$. These gradients 
correspond to the cost function $\hat{J}_{WL-TLlnC} (|\hat{e}^o(k)|)$, with respect to the weight vectors $\boldsymbol{h}(k)$ and $\boldsymbol{g}(k)$, respectively. The censoring threshold in equation \eqref{eq:46} is determined by
\begin{equation} \label{eq:47}\tag{47}
	\tau = \sqrt{\ln\left(\frac{1}{1 - P_{ce}}\right)}
\end{equation}
here, \( P_{\text{ce}} = \frac{K - p}{K} \) represents the average censoring ratio. \( K \) is the total number of data samples, \( p \) is the number of informative samples. To support real-time implementation, the noise-related parameter $\sigma_p$ can be estimated recursively as follows
\begin{equation}
    {\sigma_p}^2 = \rho {\sigma_p}^2 + (1-\rho)|\hat{e}^o(k)|^2 \tag{48}
\end{equation}
where $\rho$ is the forgetting factor constrained (0,1).

\section{Performance Analysis}
This section presents a performance analysis of the OC-WL-TLlnC method, which uses online censoring to discard less informative data or outliers. The update probability in this context is defined as
\begin{equation}
\begin{aligned}
    \text{Pr}\{c(k) = 0, c_o(k) = 0\} = & \text{Pr}\{|\hat{e}^o(k)| \geq \tau\sigma_p\} - \text{Pr}\{|\hat{e}^o(k)| > \tau^o\sigma_p\} \\
    & = 1 - P_{ce}(k)
\end{aligned}
    \tag{49}
\label{eq:49}
\end{equation}
where $c(k)$ and $c_o(k)$ represent the binary censoring variable and the additional censoring parameter. The censored update equation is given by
\begin{equation}
\begin{aligned}
    \boldsymbol{w}_o(k + 1) = & \boldsymbol{w}_o(k) + \mu (1 - P_{ce}) \tanh\left( \lambda |\hat{e}^o(k)|\ \right) \\
    & \cdot\frac{\hat{e}^o(k)}{|\hat{e}^{o}(k)|}  \cdot \left[ \frac{\boldsymbol{v}^*_o(k)}{\|\bar{\boldsymbol{w}}_{o,\dagger}\|} + \frac{\hat{e}^*(k)\boldsymbol{w}_o(k)}{\|\bar{\boldsymbol{w}}_{o,\dagger}\|^3} \right]
\end{aligned}
    \tag{50}
\label{eq:50}
\end{equation}
where $\bar{\boldsymbol{w}}_{o,\dagger} = [\boldsymbol{h}, \boldsymbol{g}, \sqrt{\gamma}]^T$, $\boldsymbol{v}^*_o(k) = [\boldsymbol{v}^*(k), \boldsymbol{v}(k)]^T$, and $\hat{e}^*(k) = v^*(k + 1) - \boldsymbol{w}^*_o(k)\boldsymbol{v}^*_o(k)$.

To begin with the performance analysis, the following assumptions are provided:
\begin{itemize}
    \item \textbf{Assumption 1:} Both the input noise $\boldsymbol{q}^*(k)$ and the output noise $p^*(k)$ are mutually independent and follow a joint zero-mean Gaussian distribution.
    \item \textbf{Assumption 2:} The augmented voltage vector $\boldsymbol{v}^*_o(k) = [\boldsymbol{v}^*(k), \boldsymbol{v}(k)]^T$ is statistically independent of both the input noise $\boldsymbol{q^*}(k)$ and the output noise $p^*(k)$. Additionally, all elements of the voltage vector are independent and identically distributed.
    \item \textbf{Assumption 3:} The voltage signal $\boldsymbol{v}(k)$ does not depend on the adaptive weight vectors $\boldsymbol{h}(k)$ and $\boldsymbol{g}(k)$.
    \item \textbf{Assumption 4:} The augmented input $\boldsymbol{v}_o(k) \in \mathbb{C}^{2L \times 1}$ has a Hermitian covariance matrix $\boldsymbol{R}_{\boldsymbol{v}_o} = \mathbb{E}[\boldsymbol{v}_o(k)\boldsymbol{v}_o^{H}(k)]$, which is positive definite and therefore full rank. In particular, $\operatorname{Tr}(\boldsymbol{R}_{\boldsymbol{v}_o}) = \mathbb{E}[\boldsymbol{v}_o^{H}(k)\boldsymbol{v}_o(k)] > 0$.
    \item \textbf{Assumption 5:} The censoring probability $P_{ce}(k) = \text{Pr}\{|\hat{e}^o(k)| > \tau\sigma_p\}$ is small, and the nonlinear function $\tanh(\lambda |\hat{e}^o(k)|)$ operates independently of the augmented voltage signal $\boldsymbol{v}^*_o(k)$ for mean convergence analysis.
\end{itemize}

\subsection{Local Stability and Convergence Analysis}
This subsection analysis the local stability of the OC-WL-TLlnC algorithm near its optimal point. Defining the augmented optimal weight vector as $\boldsymbol{w}_{o,\dagger} = [\boldsymbol{h}_{\dagger}, \boldsymbol{g}_{\dagger}]^T$, the output error at this optimal point is given by
\begin{equation}
\begin{aligned}
    \epsilon_{\dagger}(k) = \boldsymbol{w}_{o,\dagger}\boldsymbol{v}_o(k) + p(k) - \boldsymbol{w}_{o,\dagger}(\boldsymbol{v}_o(k) + \boldsymbol{l}(k))  = p(k) - \boldsymbol{w}_{o,\dagger}\boldsymbol{l}(k)
\end{aligned}
    \tag{51}
\label{eq:51}
\end{equation}
where $\boldsymbol{l}(k) = [\boldsymbol{q}(k), \boldsymbol{q}^*(k)]^T$ representing the augmented input noise vector. By multiplying both sides of \eqref{eq:51} by their conjugates and taking the expectation, and using Assumptions 1 and 2, the result is expressed as
\begin{equation}
\begin{aligned}
    E[|\epsilon_{\dagger}(k)|^2] = & E[|p(k)|^2] + E[|\boldsymbol{w}_{o,\dagger}\boldsymbol{l}(k)|^2] = \sigma^2_q \|\boldsymbol{\bar{w}}_{o,\dagger}\|^2
\end{aligned}
    \tag{52}
\label{eq:52}
\end{equation}

Following Assumptions 1 and 2, the cross-correlation $\boldsymbol{v}^*_o(k)$ and $\epsilon_{\dagger}(k)$ is determined as
\begin{equation}
\begin{aligned}
    E[\boldsymbol{v}^*_o(k)\epsilon_{\dagger}(k)] = E[\boldsymbol{l}^*(k)\epsilon_{\dagger}(k)] = -\sigma^2_q \boldsymbol{w}_{o,\dagger}
\end{aligned}
    \tag{53}
\label{eq:53}
\end{equation}

Based on the gradient of the OC-WL-TLlnC cost function with censoring, the update direction at $\boldsymbol{w}_{o,\dagger}$ is given by
\begin{equation}
\begin{aligned}
g_{TLlnC}(\boldsymbol{w}_{o,\dagger}) 
= 
\tanh\!\left( \lambda |\hat{\epsilon}_{\dagger}(k)| \right)\frac{1}{|\hat{\epsilon}_{\dagger}(k)|}
\Biggl[-
    \frac{\|\bar{\boldsymbol{w}}\|^{2}\boldsymbol{v}^*_o(k)\epsilon_{\dagger}(k)}{\|\bar{\boldsymbol{w}}\|^{4}} \\
    -\frac{|\epsilon_{\dagger}(k)|^{2}\,\boldsymbol{w}_{o,\dagger}}{\|\bar{\boldsymbol{w}}\|^{4}}
\Biggr]
\end{aligned}
\label{eq:54}
\tag{54}
\end{equation}

Using \eqref{eq:52} and \eqref{eq:53} in \eqref{eq:54} and applying the expectation operator result in
\begin{equation}
    E[g_{TLlnC}(\boldsymbol{w}_o(k))] = 0
    \tag{55}
\label{eq:55}
\end{equation}

Hence, $E[g_{TLlnC}(\boldsymbol{w}_{o,\dagger})]$ is identified as a critical point of the cost function. To further validate the analysis, we next compute the augmented Hessian matrix $\boldsymbol{H}(\boldsymbol{w}_o(k))$, which is defined as \cite{22,36}
\begin{equation}
    \boldsymbol{H}(\boldsymbol{w}_o(k)) = \begin{bmatrix}
        \boldsymbol{H}_1(\boldsymbol{w}_o(k)) & \boldsymbol{H}_2(\boldsymbol{w}_o(k)) \\
        \boldsymbol{H}^*_2(\boldsymbol{w}_o(k)) & \boldsymbol{H}^*_1(\boldsymbol{w}_o(k))
    \end{bmatrix}
    \tag{56}
\label{eq:56}
\end{equation}

The individual components of the Hessian matrix are defined as $\boldsymbol{H}_1 (\boldsymbol{w}_o (k)) = \frac{\partial g_{TLlnC}(\boldsymbol{w}_o(k))}{\partial \boldsymbol{w}^T}$, $\boldsymbol{H}_2 (\boldsymbol{w}_o (k)) = \frac{\partial g_{TLlnC}(\boldsymbol{w}_o(k))}{\partial \boldsymbol{w}^H}$, and their corresponding expressions are given as follows
\begin{equation}
\begin{aligned}
\boldsymbol{H}_1
&=
\operatorname{sech}^{2}\!\big(\lambda|\hat{e}^o(k)|\big)\,
\Bigg[
-\frac{\|\bar{\boldsymbol{w}}\|^{2}\,\boldsymbol{v}_o^{T}(k)\,\hat{e}^*(k)}{\|\bar{\boldsymbol{w}}\|^{4}}
-\frac{|\hat{e}(k)|^{2}}{\|\bar{\boldsymbol{w}}\|^{4}}\,\boldsymbol{w}_o^{H}(k)
\Bigg]
\\
&\quad\,
\frac{1}{|\hat{e}^o(k)|^{2}}\cdot
\Bigg[
-\frac{\|\bar{\boldsymbol{w}}\|^{2}\,\boldsymbol{v}^*_o(k)\,\hat{e}(k)}{\|\bar{\boldsymbol{w}}\|^{4}}
-\frac{|\hat{e}(k)|^{2}}{\|\bar{\boldsymbol{w}}\|^{4}}\,\boldsymbol{w}_o(k)
\Bigg]
\\
&\quad
+
\operatorname{tanh}\!\big(\lambda|\hat{e}^o(k)|\big)\, \frac{1}{|\hat{e}^o(k)|^{3}}\,
\Bigg[
\frac{\|\bar{\boldsymbol{w}}\|^{2}\,\boldsymbol{v}_o^{T}(k)\,\hat{e}^*(k)}{\|\bar{\boldsymbol{w}}\|^{4}}
+\frac{|\hat{e}(k)|^{2}}{\|\bar{\boldsymbol{w}}\|^{4}}\,\boldsymbol{w}_o^{H}(k)
\Bigg]
\\
&\quad\cdot
\Bigg[
-\frac{\|\bar{\boldsymbol{w}}\|^{2}\,\boldsymbol{v}^*_o(k)\,\hat{e}(k)}{\|\bar{\boldsymbol{w}}\|^{4}}
-\frac{|\hat{e}(k)|^{2}}{\|\bar{\boldsymbol{w}}\|^{4}}\,\boldsymbol{w}_o(k)
\Bigg] +\Bigg(\Bigg[
\frac{\boldsymbol{v}^*_o(k)\boldsymbol{v}_o^{T}(k)}{\|\bar{\boldsymbol{w}}\|^{2}}
-\frac{|\hat{e}(k)|^{2}}{\|\bar{\boldsymbol{w}}\|^{4}} \\
&\quad
-\frac{2\,\boldsymbol{w}_o^{H}(k)\,\boldsymbol{v}^*_o(k)\,\hat{e}(k)}{\|\bar{\boldsymbol{w}}\|^{4}}
+\frac{2\,\boldsymbol{v}_o^{T}(k)\,\hat{e}^*(k)\,\boldsymbol{w}_o(k)}{\|\bar{\boldsymbol{w}}\|^{4}}
\Bigg] -\frac{4}{\|\bar{\boldsymbol{w}}\|^{2}}\,
\boldsymbol{w}_o^{H}(k) \\
&\quad
\Bigg[
-\frac{\|\bar{\boldsymbol{w}}\|^{2}\,\boldsymbol{v}^*_o(k)\,\hat{e}(k)}{\|\bar{\boldsymbol{w}}\|^{4}}
-\frac{|\hat{e}(k)|^{2}}{\|\bar{\boldsymbol{w}}\|^{4}}\,\boldsymbol{w}_o(k)
\Bigg]\Bigg)\,\operatorname{tanh}\!\big(\lambda|\hat{e}^o(k)|\big)\,
\frac{1}{|\hat{e}^o(k)|}
\end{aligned}
\label{eq:57}
\tag{57}
\end{equation}
\begin{equation}
\begin{aligned}
\boldsymbol{H}_2
&=
\operatorname{sech}^{2}\!\big(\lambda|\hat{e}^o(k)|\big)\,
\Bigg[
-\frac{\|\bar{\boldsymbol{w}}\|^{2}\,\boldsymbol{v}^H_o(k)\,\hat{e}(k)}{\|\bar{\boldsymbol{w}}\|^{4}}
-\frac{|\hat{e}(k)|^{2}}{\|\bar{\boldsymbol{w}}\|^{4}}\,\boldsymbol{w}^T_o(k)
\Bigg]\,
\frac{1}{|\hat{e}^o(k)|^{2}}
\\
&\quad\cdot
\Bigg[
-\frac{\|\bar{\boldsymbol{w}}\|^{2}\,\boldsymbol{v}_o^*(k)\,\hat{e}(k)}{\|\bar{\boldsymbol{w}}\|^{4}}
-\frac{|\hat{e}(k)|^{2}}{\|\bar{\boldsymbol{w}}\|^{4}}\,\boldsymbol{w}_o(k)
\Bigg] + \operatorname{tanh}\!\big(\lambda|\hat{e}^o(k)|\big)\,\frac{1}{|\hat{e}^o(k)|^{3}}\\[10pt]
&\quad 
 \Bigg[
\frac{\|\bar{\boldsymbol{w}}\|^2 \boldsymbol{v}^H_o(k)\,\hat{e}(k)}{\|\bar{\boldsymbol{w}}\|^4}
+ \frac{|\hat{e}(k)|^2\boldsymbol{w}^T_o(k)}{\|\bar{\boldsymbol{w}}\|^4}\Bigg]\,
\Bigg[
-\frac{\|\bar{\boldsymbol{w}}\|^{2}\,\boldsymbol{v}^*_o(k)\,\hat{e}(k)}{\|\bar{\boldsymbol{w}}\|^{4}}
-\\
&\quad \frac{|\hat{e}(k)|^{2}\boldsymbol{w}_o(k)}{\|\bar{\boldsymbol{w}}\|^{4}}
\Bigg]
+ \Biggl(\Bigg[
-\frac{2\,\boldsymbol{w}^T_o(k)\,\boldsymbol{v}^*_o(k)\,\hat{e}(k)}{\|\bar{\boldsymbol{w}}\|^{4}}
+ \frac{2\,\boldsymbol{v}^H_o(k)\,\hat{e}(k)\,\boldsymbol{w}_o(k)}{\|\bar{\boldsymbol{w}}\|^{4}}
\Bigg]\, - \\
&\quad \frac{4}{\|\bar{\boldsymbol{w}}\|^{2}}\,
\boldsymbol{w}^T_o(k)\,
\Bigg[
-\frac{\|\bar{\boldsymbol{w}}\|^{2}\,\boldsymbol{v}^*_o(k)\,\hat{e}(k)}{\|\bar{\boldsymbol{w}}\|^{4}}
-\frac{|\hat{e}(k)|^{2}}{\|\bar{\boldsymbol{w}}\|^{4}}\,\boldsymbol{w}_o(k)
\Bigg]\Biggl)\,\\& \quad \cdot \operatorname{tanh}\!\big(\lambda|\hat{e}^o(k)|\big)\,
\frac{1}{|\hat{e}^o(k)|}
\end{aligned}
\label{eq:58}
\tag{58}
\end{equation}

Next, the augmented Hessian matrix $\boldsymbol{H}(\boldsymbol{w}_o (k))$ is examined at $\boldsymbol{w}_{o,\dagger}$. By substituting \eqref{eq:52} and \eqref{eq:53} into \eqref{eq:57} and \eqref{eq:58} and replacing $\boldsymbol{w}_o (k)$ with $\boldsymbol{w}_{o,\dagger}$, and applying Assumptions 1-5, the resulting expression is obtained as follows
\begin{equation}
\begin{aligned}
E[\boldsymbol{H}_1(\boldsymbol{w}_{o,\dagger})] = \left( \frac{\mathrm{Tr}\boldsymbol{(\mathbf{R})}}{\|\bar{\boldsymbol{w}}_{o,\dagger}\|^2}
\right) \cdot \frac{E[\tanh\!\left(\lambda |\hat{\epsilon}_{\dagger}(k)|\right)]}{\sqrt{\sigma^2_q}}
\end{aligned}
\tag{59}\label{eq:59}
\end{equation}

In order to simplify the nonlinear function $\tanh(\lambda |\hat{\epsilon}_\dagger(k)|)$, we expand $|\hat{\epsilon}_\dagger(k)|$ around zero using a Taylor series. The expansion can be written explicitly as follows
\begin{equation}
\tanh(\lambda |\hat{\epsilon}_\dagger(k)|) = \lambda |\hat{\epsilon}_\dagger(k)| - \frac{\lambda^3 |\hat{\epsilon}_\dagger(k)|^3}{3} + o(|\hat{\epsilon}_\dagger(k)|^3)\tag{60}
\end{equation}
where $o(|\hat{\epsilon}_\dagger(k)|^3)$ denotes higher-order terms. Neglecting these higher-order terms allows us to approximate the expected value as
\begin{equation}
\begin{aligned}
E[\tanh(\lambda |\hat{\epsilon}_\dagger(k)|)] &\approx E\Big[\lambda |\hat{\epsilon}_\dagger(k)| - \frac{\lambda^3 |\hat{\epsilon}_\dagger(k)|^3}{3}\Big] \\
&\approx E\Big[\frac{\lambda |\epsilon_\dagger(k)|}{\|\bar{\boldsymbol{w}}\|} - \frac{\lambda^3 |\epsilon_\dagger(k)|^3}{3 \|\bar{\boldsymbol{w}}\|^3}\Big] \\
&\approx \lambda \sqrt{\sigma_q^2} - \frac{\lambda^3 \sigma_q^2 \sqrt{\sigma_q^2}}{3}
\end{aligned}\tag{61}\label{eq:61}
\end{equation}

Substituting \eqref{eq:61} into \eqref{eq:59}, we obtain \eqref{eq:62} as follows
\begin{equation}
\begin{aligned}
E[\boldsymbol{H}_1(\boldsymbol{w}_{o,\dagger})] = 
\left( \frac{\mathrm{Tr}\boldsymbol{(\mathbf{R})}}{\|\bar{\boldsymbol{w}}_{o,\dagger}\|^2}\right)\Bigg(\lambda -\frac{\lambda^3 \sigma_q^2}{3}\Bigg)
\end{aligned}
\tag{62}\label{eq:62}
\end{equation}

Following the same procedure, we substitute \eqref{eq:52} and \eqref{eq:53} into \eqref{eq:58}, leading to
\begin{equation}
\boldsymbol{H}_2(\boldsymbol{w}_{o,\dagger}) = 0
\tag{63}
\label{eq:63}
\end{equation}

By inserting \eqref{eq:62} and \eqref{eq:63} into \eqref{eq:56}, the Hessian matrix at $\boldsymbol{w}_{o,\dagger}$ becomes
\begin{equation}
\begin{aligned}
\mathbf{H}(\boldsymbol{w}_{o\dagger}) 
&= 
\frac{\Big(3\lambda -\lambda^3 \sigma_q^2 \Big)}
     {3\, \|\bar{\boldsymbol{w}}_{o\dagger}\|^2} 
\begin{bmatrix} 
\boldsymbol{\mathrm{Tr}\boldsymbol{(\mathbf{R})}} & \mathbf{0} \\ 
\mathbf{0} & \boldsymbol{\mathrm{Tr}\boldsymbol{(\mathbf{R})}}^* 
\end{bmatrix}
\end{aligned}
\tag{64}\label{eq:64}
\end{equation}

This positive definite Hessian matrix confirms the local stability of the OC-WL-TLlnC algorithm at the optimal solution.
\subsection{Convergence Condition and Step-size Bound}
Based on Assumption 4, $\boldsymbol{H}(\boldsymbol{w}_{o,\dagger})$ is confirmed to be positive definite. Together with \eqref{eq:55}, this shows that $\boldsymbol{w}_{o,\dagger}$ is indeed a local minimum point for the cost function \eqref{eq:42}. We now examine the step size $\mu$ required by the proposed OC-WL-TLlnC method to ensure convergence. We introduce the augmented weight error vector $\Delta\boldsymbol{w}_{\circ}(k)$ as the difference between $\boldsymbol{w}_{o,\dagger} - \boldsymbol{w}_{\circ}(k)$. Using this, the recursion for $\Delta\boldsymbol{w}_{\circ}(k)$ can be obtained as follows
\begin{equation}
\Delta\boldsymbol{w}_{\circ}(k + 1) 
= 
\Delta\boldsymbol{w}_{\circ}(k) 
+ \mu (1-P_{ce}) g_{TLlnC}(\boldsymbol{w}_{\circ}(k))
\tag{65}\label{eq:65}
\end{equation}

To move forward, we use a Taylor series expansion to approximate $g_{TLlnC}(\boldsymbol{w}_{\circ}(k))$. First, near the best solution $\boldsymbol{w}_{o,\dagger}$, the weight error $\Delta\boldsymbol{w}_{\circ}(k)$ is small. Secondly, as shown in Assumption 5, the nonlinear component of the update can be considered independently of the input signal for our analysis. This results in the subsequent approximation
\begin{equation}
\begin{aligned}
    g_{TLlnC}(\boldsymbol{w}_{\circ}(k)) \approx & g_{TLlnC}(\boldsymbol{w}_{o,\dagger}) - \boldsymbol{H}_1(\boldsymbol{w}_{o,\dagger}) \Delta\boldsymbol{w}_{\circ}(k) \\
    & - \boldsymbol{H}_2(\boldsymbol{w}_{o,\dagger}) \Delta\boldsymbol{w}_{\circ}^*(k)
\end{aligned}
    \tag{66}
\label{eq:66}
\end{equation}

Given the minimal weight error $\Delta\boldsymbol{w}_{\circ}(k)$ near the optimal solution, we can streamline the formula. It also established from \eqref{eq:63} that $\boldsymbol{H}_2(\boldsymbol{w}_{o,\dagger})$ equals zero. Utilizing these facts in \eqref{eq:66} yields a more simplified expression
\begin{equation}
    g_{TLlnC}(\boldsymbol{w}_{\circ}(k)) \approx- \boldsymbol{H}_1(\boldsymbol{w}_{o,\dagger}) \Delta\boldsymbol{w}_{\circ}(k)
    \tag{67}
\label{eq:67}
\end{equation}

We now substitute this key result from \eqref{eq:67} directly into the weight-error update rule in \eqref{eq:68}. To analyze the average convergence behavior, we take the expectations, which leads to the following dynamic equation for the mean weight error as
\begin{equation}
\begin{aligned}
    E[\Delta\boldsymbol{w}_{\circ}(k + 1)] =  \left[ \boldsymbol{I} - \mu (1-P_{ce}) \boldsymbol{H}_1(\boldsymbol{w}_{o,\dagger}) \right] 
     \cdot E[\Delta\boldsymbol{w}_{\circ}(k)]
\end{aligned}
    \tag{68}
\label{eq:68}
\end{equation}

Based on \eqref{eq:68}, to guarantee local convergence in mean sense, we need to ensure all eigenvalues of $\boldsymbol{I} - \mu (1-P_{ce}) \boldsymbol{H}_1(\boldsymbol{w}_{o,\dagger})$ lie inside the unit circle. This leads to the following condition
\begin{equation}
    \| \boldsymbol{I} - \mu (1-P_{ce}) \boldsymbol{H}_1(\boldsymbol{w}_{o,\dagger}) \| < 1
    \tag{69}
\label{eq:69}
\end{equation}

Ultimately, we replace the outcome for \eqref{eq:62} into \eqref{eq:69}, we obtain the convergence condition for the proposed OC-WL-TLlnC algorithm as
\begin{equation}
0 < \mu < 
\frac{6 \, \|\bar{\boldsymbol{w}}_{o\dagger}\|^2}
     {(1 - P_{ce}) \, \Big(3\lambda -\lambda^3 \sigma_q^2\Big) \, \lambda_{\max}(\mathrm{Tr}\boldsymbol{(\mathbf{R})})}
\tag{70}\label{eq:70}
\end{equation}
where $\lambda_{\text{max}}(\mathrm{Tr}\boldsymbol{(\mathbf{R})})$ denotes the maximum eigenvalue of the matrix $\mathrm{Tr}\boldsymbol{(\mathbf{R})}$.

This convergence condition ensures that the proposed algorithm converges to the optimal solution when the step size $\mu$ satisfies the bound in \eqref{eq:70}. The bound depends on the size of the optimal weight vector, the input signal covariance matrix, the input noise power, the censoring probability, and the nonlinear characteristics of the cost function.

\subsection{Steady-state Mean Square Behavior}
The gradient error is defined as $\boldsymbol{e}(k) \triangleq \hat{g}_{TLlnC}(\boldsymbol{w}_{\circ}(k)) - g_{TLlnC}(\boldsymbol{w}_{\circ}(k))$, which denotes the difference between the expected and instantaneous gradients. The update equation is therefore given by 
\begin{equation}
    \Delta\boldsymbol{w}_{\circ}(k + 1) = \Delta\boldsymbol{w}_{\circ}(k) - \mu (1-P_{ce}) g_{TLlnC}(\boldsymbol{w}_{\circ}(k)+\boldsymbol{e}(k))
    \tag{71}
\label{eq:71}
\end{equation}

In steady-state conditions, $\boldsymbol{w}_{\circ}(k)$ converges to $\boldsymbol{w}_{o,\dagger}$, leading to $\boldsymbol{e}(\boldsymbol{w}_{o,\dagger}) = \hat{g}_{TLlnC}(\boldsymbol{w}_{o,\dagger})$. Consequently, \eqref{eq:71} reduces to
\begin{equation}
    \Delta\boldsymbol{w}_{\circ}(k + 1) = \Delta\boldsymbol{w}_{\circ}(k) - \mu (1-P_{ce}) \hat{g}_{TLlnC}(\boldsymbol{w}_{o,\dagger})
    \tag{72}
\label{eq:72}
\end{equation}

The expected value of the weighted squared norm of the augmented weight error vector, $\|\Delta\boldsymbol{w}_{\circ}(k)\|^2_{\boldsymbol{M}} = \Delta\boldsymbol{w}_{\circ}^H(k)\boldsymbol{M}\Delta\boldsymbol{w}_{\circ}(k)$, where $\boldsymbol{M} \in \mathbb{C}^{2L \times 2L}$ is a semi-definite weight matrix, describes the MSD of the proposed OC-WL-TLlnC method. Based on Assumptions 2 and 3, the recursion can be formulated from \eqref{eq:72} as follows 
\begin{equation}
\begin{aligned}
E[\|\Delta\boldsymbol{w}_{\circ}(k + 1)\|^2_{\boldsymbol{M}}] =  E[\|\Delta\boldsymbol{w}_{\circ}(k)\|^2_{\boldsymbol{U}}]  + \mu^2  (1-P_{ce})^2 E[\|\hat{g}_{TLlnC}(\boldsymbol{w}_{o,\dagger})\|^2_{\boldsymbol{M}}]
\end{aligned}
    \tag{73}
\label{eq:73}
\end{equation}
where
\begin{equation}
\begin{aligned}
    \boldsymbol{U} =  \left( \boldsymbol{I} - \mu (1-P_{ce}) \boldsymbol{H}_1(\boldsymbol{w}_{o,\dagger}) \right)^H  \cdot \boldsymbol{M} \left( \boldsymbol{I} - \mu (1-P_{ce}) \boldsymbol{H}_1(\boldsymbol{w}_{o,\dagger}) \right)
\end{aligned}
    \tag{74}
\label{eq:74}
\end{equation}

Let $\boldsymbol{\zeta} \triangleq \text{vec}(\boldsymbol{M})$ represent the vectorization of the weight matrix. Then, \eqref{eq:74}, leads to 
\begin{equation} \label{eq:75} 
\begin{aligned}
\boldsymbol{G}\boldsymbol{\zeta} &\triangleq \text{vec}(\boldsymbol{U})
\end{aligned}\tag{75}
\end{equation}
here, $\boldsymbol{G}$ is given by $\boldsymbol{G} =
\left( \boldsymbol{I} - \mu (1-P_{ce}) \boldsymbol{H}_1(\boldsymbol{w}_{o,\dagger}) \right)^T
\otimes$
$\left( \boldsymbol{I} - \mu (1-P_{ce}) \boldsymbol{H}_1(\boldsymbol{w}_{o,\dagger}) \right)^{H}.$\\
With this definition, the recursion in \eqref{eq:73} becomes
\begin{equation}
\begin{aligned}
    E[\|\Delta\boldsymbol{w}_{\circ}(k + 1)\|^2_{\text{vec}^{-1}(\boldsymbol{\zeta})}] = & E[\|\Delta\boldsymbol{w}_{\circ}(k)\|^2_{\text{vec}^{-1}(\boldsymbol{G}\boldsymbol{\zeta})}] \\
    & + \mu^2(1-P_{ce})^2 (\text{vec}(\boldsymbol{S}))^H \boldsymbol{\zeta}
\end{aligned}
    \tag{76}
\label{eq:76}
\end{equation}
where
\begin{equation}
    \boldsymbol{S} = E[\hat{g}_{TLlnC}(\boldsymbol{w}_{o,\dagger}) \hat{g}_{TLlnC}^H(\boldsymbol{w}_{o,\dagger})]
    \tag{77}
\label{eq:77}
\end{equation}

Explicitly, $\boldsymbol{S}$ can be written using the statistical properties from \eqref{eq:52} and \eqref{eq:53} and the gradient expression
\begin{equation}
\begin{aligned}
\boldsymbol{S} = &E[\tanh^{2} (\lambda|\hat{\epsilon}_{\dagger}(k)|)] \cdot \\
&\left[\frac{\left(8L^{2}\sigma_{q}^{2}\|\bar{\boldsymbol{w}}_{o,\dagger}\|+3\sigma_{q}^{2}\|\bar{\boldsymbol{w}}_{o,\dagger}\|-24L^{2}\sigma_{q}^{2}\boldsymbol{w}_{o,\dagger}\boldsymbol{w}^H_{o,\dagger}-4\sigma_{p}^{2}\right)\boldsymbol{w}^H_{o,\dagger}\boldsymbol{w}_{o,\dagger}}{\|\bar{\boldsymbol{w}}_{o,\dagger}\|^{6}}\right.\\
&+\left.\frac{\left(\sigma_{p}^{2}+2L\sigma_{q}^{2}\boldsymbol{w}_{o,\dagger}\boldsymbol{w}^H_{o,\dagger}\right)\mathbf{I}}{\|\bar{\boldsymbol{w}}_{o,\dagger}\|^{4}}+\frac{\mathrm{Tr}\boldsymbol{(\mathbf{R}_{v_o})}}{\|\bar{\boldsymbol{w}}_{o,\dagger}\|^2}\right]
\end{aligned}
\label{eq:78}
\tag{78}
\end{equation}
where $\mathrm{Tr}\boldsymbol{(\mathbf{R}_{v_o})}=\mathrm{Tr}\boldsymbol{(\mathbf{R})}+\sigma^2_P\boldsymbol{I}$. Because the step-size condition in \eqref{eq:70} is satisfied, $\boldsymbol{G}$ is stable, which ensures the recursion in \eqref{eq:76} is mean-square stable. Therefore, the MSD at steady-state is given by
\begin{equation}
\begin{aligned}
    E[\|\Delta\boldsymbol{w}_{\circ}(\infty)\|^2] =  \mu^2 (1-P_{ce})^2 (\text{vec}(\boldsymbol{S}))^H  \cdot (\boldsymbol{I} - \boldsymbol{G})^{-1} \text{vec}(\boldsymbol{I})
\end{aligned}
    \tag{79}
\label{eq:79}
\end{equation}

This expression provides the steady-state MSD for the proposed OC-WL-TLlnC method. It shows the trade-off between convergence speed (controlled by $\mu$) and steady-state accuracy. The effect of censoring is included through the $(1 - P_{ce})$ factors. The performance also depends on the input signal statistics $\mathrm{Tr}\boldsymbol{(\mathbf{R}_{v_o})}$, the noise variances $\sigma_p^2$ and $\sigma_q^2$, the nonlinear characteristics through the $E[\tanh^2(\cdot)]$ term, and the properties of the optimal weight vector.

    \section{Simulation results}
\label{sec6}
Numerical simulations were conducted in MATLAB to evaluate the OC-WL-TLlnC algorithm under various unbalanced conditions in a three-phase power system. Comparative analyses were also carried out using ACLlncosh, ACMEE, ACLMS, OC-CTLlnC, CLlncosh, CMEE and CLMS based on the simulation outcomes. Unless otherwise specified, the system frequency was 50 Hz, and the sampling rate was 2.5 kHz. For specific scenarios, a sampling rate of 5kHz was used to better evaluate performance. A type D voltage sag, as described in \cite{29}, was used to introduce an imbalance into the system. Background noise followed a Gaussian distribution, and the signal-to-noise ratio (SNR) was varied to assess the algorithm’s performance under different noise conditions. Impulsive interference was modeled using a Bernoulli-Gaussian process activated with probability $p = 0.005$ and variance $\sigma_{impulsive}^2 = 100\sigma^2$. The algorithms were initialized with the following parameters: (i) OC-WL-TLlnC and OC-CTLlnC: $P_{ce} = 35\%$, $\tau^o = 2$, $\mu = 0.045$, $\lambda(k) = 1$, and $\beta = 0.9999$; (ii) ACLlncosh and CLlncosh: $\mu = 0.02$, $\lambda(k) = 1$, and $\beta = 0.9999$; (iii) CMEE and ACMEE: $\mu = 0.03$, $L = 10$, and $\sigma = 0.1$; (iv) ACLMS and CLMS: $\mu = 0.03$.

\subsection{Assessment of methods for dynamic tracking}\label{subsec1}
In the first experiment phase, the proposed methods were tested under unbalanced conditions. A type D voltage sag occurred at t=0.64s,  causing a 6.6\% drop and $8^\circ$ phase shift in two phases and a 30\% amplitude loss in the third. At t=1.73 s, a full (100\%) voltage sag was applied to phase C, further disturbing the balance. 
\begin{figure}[htb!]
    \centering
    \includegraphics[width=0.75\textwidth]{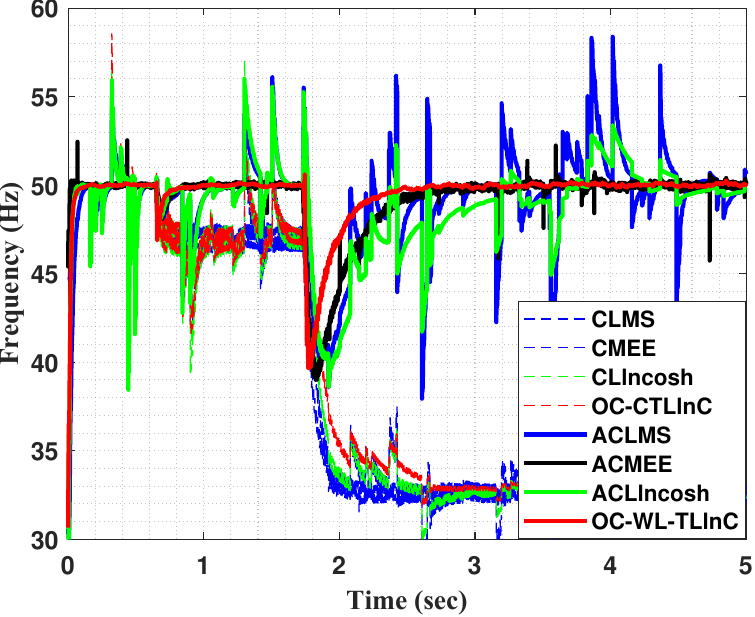}
    \caption{Performance comparison of frequency tracking methods.}
    \label{fig_enter-label1}
\end{figure}
As shown in Fig. \ref{fig_enter-label1}, the OC-CTLlnC method, constrained by a linear model, failed to track frequency accurately, leading to oscillations. This is due to its inability to capture frequency variations under non-circular, unbalanced conditions. In contrast, OC-WL-TLlnC, ACLlncosh, ACMEE, and ACLMS maintained accurate tracking, owing to their suitability for non-circular signal processing. Among them, OC-WL-TLlnC achieved the most accurate and stable frequency tracking performance.

\subsection{Performance evaluation under steady-state and varying SNRs}\label{subsec2}
The robustness of the proposed OC-WL-TLlnC method was evaluated by examining its frequency estimation variance under different SNR levels. The simulation setup introduced unbalanced system conditions through type D voltage sags, along with impulsive noise and noisy input-output generated using a Bernoulli-Gaussian model. 
\begin{figure}[htb!]
    \centering
    \includegraphics[width=0.75\textwidth]{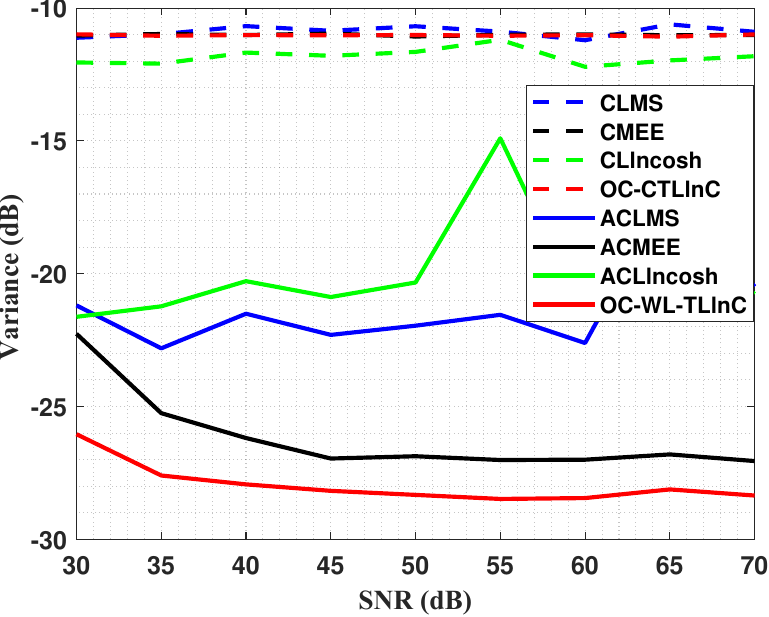}
    \caption{Comparison of frequency estimation variance under steady-state and variance SNRs.}
    \label{fig_enter-label2}
\end{figure}
Fig. \ref{fig_enter-label2}
presents the variance results obtained from 100 independent trials. The OC-WL-TLlnC algorithm maintained a consistently low estimation variance, even under harsh noise conditions. This improvement is more pronounced at lower SNR levels, where conventional algorithms tend to show significant degradation in estimation performance.   
  
\subsection{Experimental assessment}\label{subsec3}
\begin{figure}[htb!]
    \centering
    \includegraphics[width=0.75\textwidth]{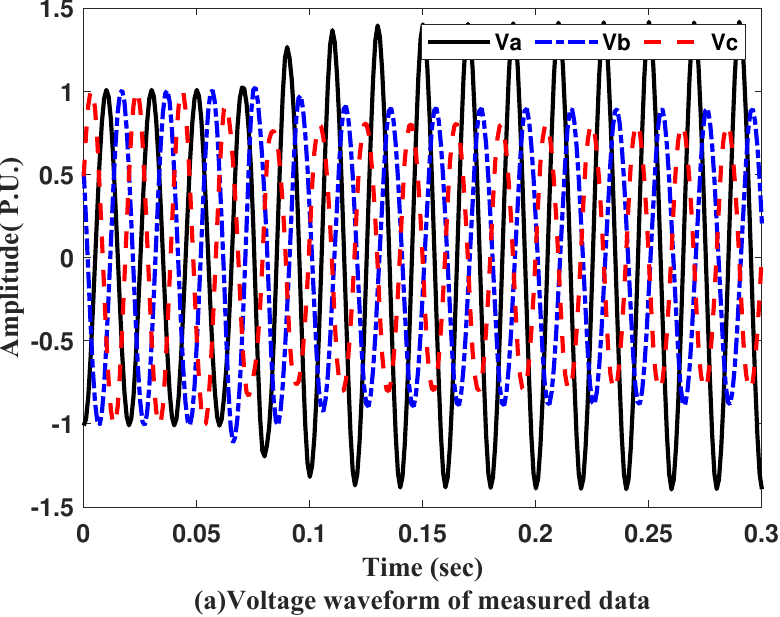}
    \includegraphics[width=0.75\textwidth]{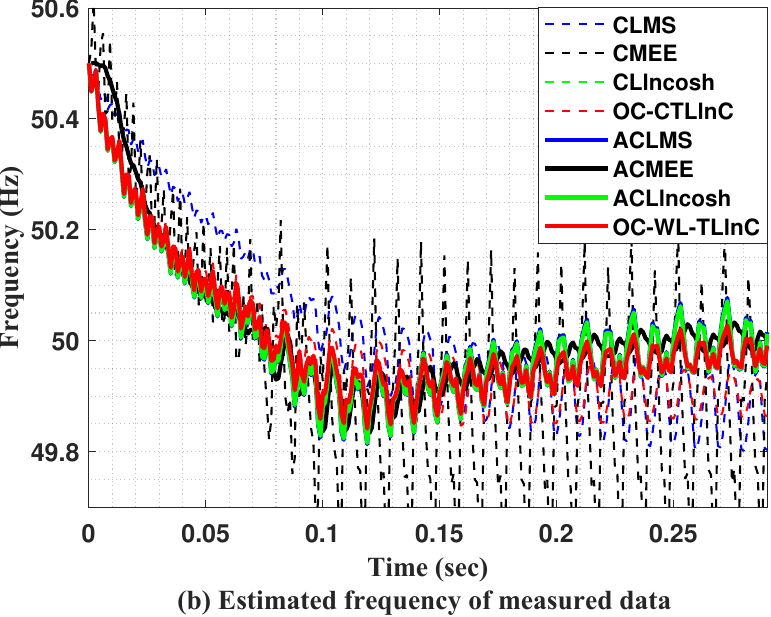}
    \caption{(a) Real-world three-phase voltage data demonstrating unbalance over time, and (b) Frequency estimation performance under real-world voltage imbalance conditions.}
    \label{Fig3}
\end{figure}
This study further examines the performance of the OC-WL-TLlnC method using real-world measurement data collected during a grounding fault at a 110/20/10 kV substation. For simulation purposes, the data were sampled at 5 kHz. The incident triggered a type D voltage sag, and the corresponding three-phase voltage waveforms are depicted in Fig. \ref{Fig3}(a). During the disturbance, voltage magnitudes in phases B and C declined by 11\% and 20\%, respectively, while phase A showed a notable increase of 41\%. As illustrated in Fig. \ref{Fig3}(b), frequency estimates produced by standard linear-based methods, including CLMS CMEE, CLlncosh, and OC-CTLlnC diverge significantly from the true system frequency, indicating their limitations under unbalanced fault conditions. 
By comparison, the proposed OC-WL-TLlnC method achieves faster convergence, reduced fluctuations, and superior tracking accuracy, confirming its robustness in distorted and asymmetrical power system environments.

\subsection{Harmonics Impact}\label{subsec4}
\begin{figure}[htb!]
    \centering
    \includegraphics[width=0.75\textwidth]{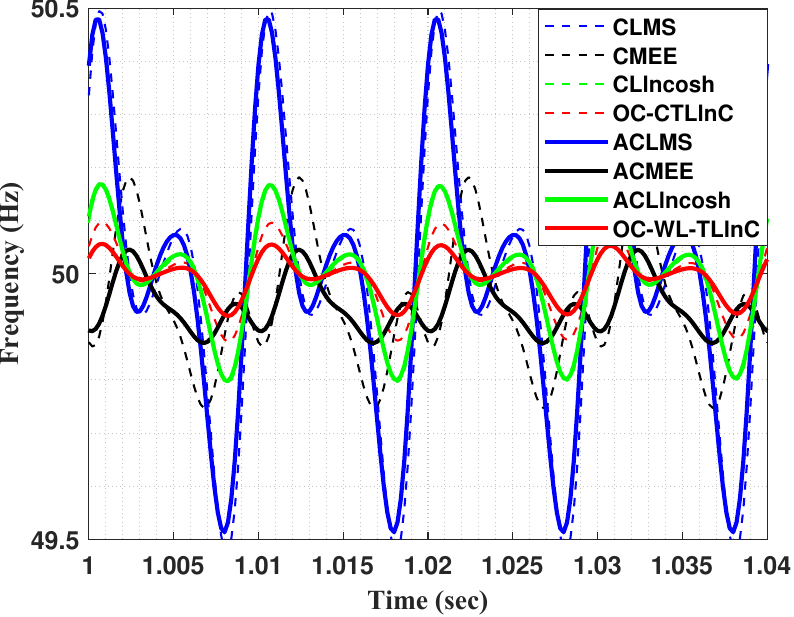}
    \caption{The steady-state frequency estimation results when higher-order
harmonics are presence.}
    \label{Fig4}
\end{figure}
The simulation in this section examines the performance of the proposed
method in the presence of harmonics. Since harmonics are common in power systems, the frequency estimator is tested using a balanced three-phase voltage signal with only measurement errors. The signal is sampled at 5 kHz. At t=0.66s, the voltage waveform contains a 7\% fifth-harmonic component and a 15\% third-harmonic component. As shown in Fig. \ref{Fig4}, the frequency estimate will exhibit oscillatory inaccuracy in the steady state when higher harmonics are present in the three-phase voltage signal. All four algorithms share oscillations ranging from around 0.05 to 0.4 Hz. The proposed approach outperforms the CLMS, CMEE, CLlncosh, OC-CTLlnC, ACLMS, ACMEE, and ACLlncosh algorithms, demonstrating superior steady-state performance even under harmonic settings. This is because the proposed algorithm accounts for the PDF error function of the system’s frequency estimate, variations in the negative-sequence component of the three-phase voltages, and the effects of harmonics and noise.

\subsection{Balanced System}\label{subsec5}
\begin{figure}[htb!]
    \centering
    \includegraphics[width=0.75\textwidth]{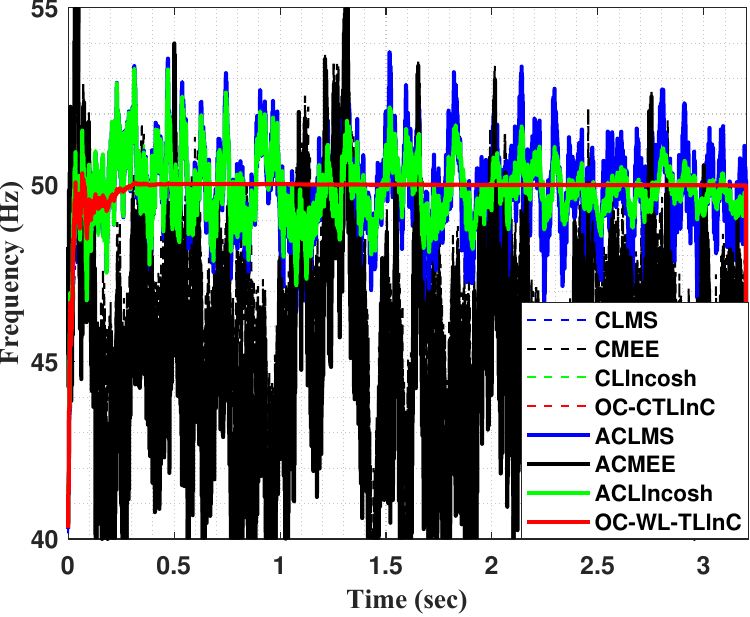}
    \caption{Performance of frequency estimation algorithms under combined 30dB measurement noise and impulsive noise in balanced three-phase voltage signals.}
    \label{Fig5}
\end{figure}
This work examined the resistance of the OC-WL-TLlnC method against measurement noise in a balanced three-phase voltage system operating at 50Hz. The system was affected by zero-mean complex-valued double white Gaussian noise with a signal-to-noise ratio of 30dB and by impulsive noise with a specified probability of occurrence. The initial frequency for all methods was 40 Hz. As shown in Fig. \ref{Fig5}, the OC-WL-TLlnC algorithm achieved accurate and stable frequency estimation even under impulsive noise. In contrast, the CLMS, CMEE, CLlncosh, ACLMS, ACMEE, ACLlncosh methods showed weaker performance. This result supports the view that the OC-WL-TLlnC method offers greater robustness under non-Gaussian noise and EIV model conditions, where traditional methods relying on second-order statistics often lose accuracy. 

\subsection{Influence of Osculatory Voltage Fluctuations on Frequency Estimation}\label{subsec6}
\begin{figure}[htb!]
    \centering
    \includegraphics[width=0.75\textwidth]{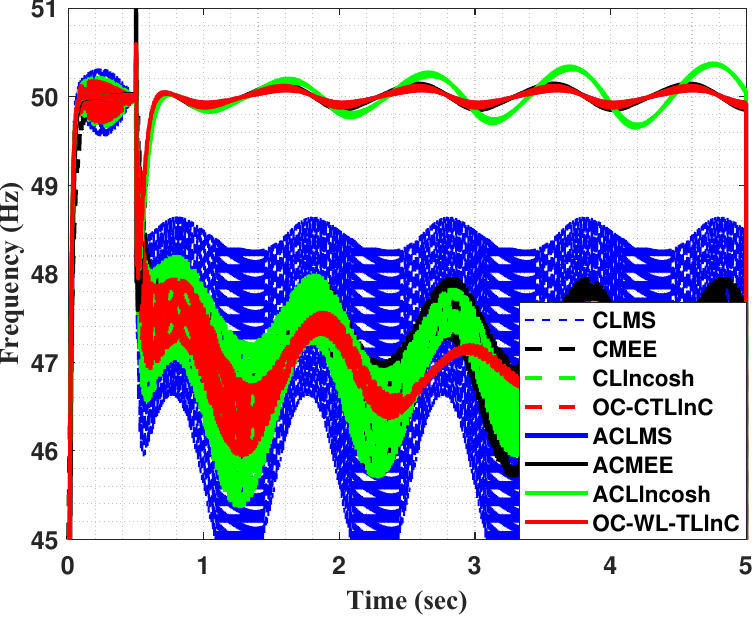}
    \caption{Effect of voltage oscillations on frequency estimation accuracy for different methods.}
    \label{Fig6}
\end{figure}
In this test, the methods were compared when the voltage amplitude changed at 1 Hz. The phase voltages were set as $v_a(k) = 1 + 0.05\sin(2\pi k T)$, 
$v_b(k) = 1 + 0.1\sin(2\pi k T)$, and $v_c(k) = 1 + 0.1\sin(2\pi k T)$, where the sampling frequency was 2 kHz. The imbalance started at $t = 0.5$s. Fig. \ref{Fig6} shows that OC-WL-TLlnC gave an accurate and steady frequency estimate with small error. The ACLlncosh, ACMEE, and ACLMS methods yielded estimated frequencies of about 0.27 Hz. CLMS, CMEE, CLlncosh, and OC-CTLlnC assume that the voltage amplitude remains constant between two samples ($A(k+1) \approx A(k)$). This assumption leads to errors when the amplitude changes, thereby reducing the methods’ reliability.

\subsection{Steady-State MSD Verification}\label{subsec7}
\begin{figure}[t!]
    \centering
    \includegraphics[width=0.75\textwidth]{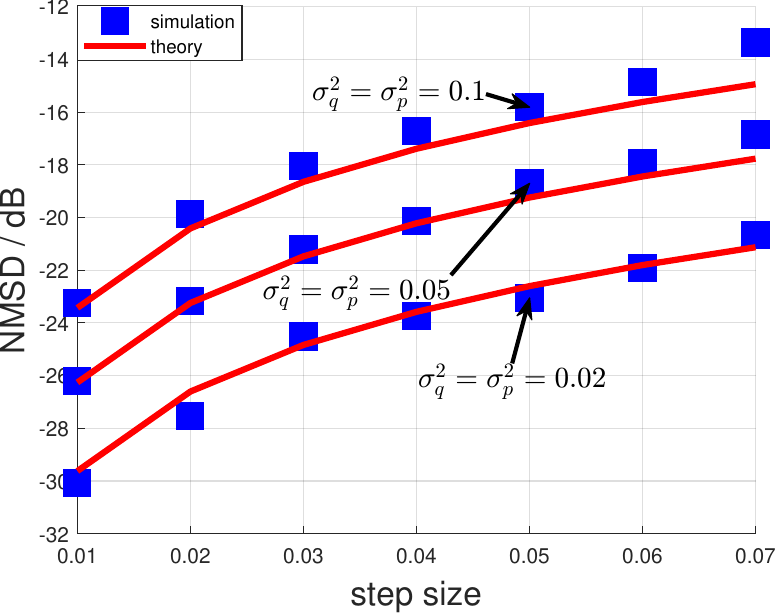}
    \caption{Comparison between theoretical and simulated steady-state MSD of the OC-WL-TLlnC algorithm under zero-mean Gaussian input and output noises with $\sigma_q^2 = 0.1$ and $\sigma_p^2 = 0.1$.}
    \label{Fig7}
\end{figure}

This section validates the theoretical steady-state mean-squared deviation (MSD) of the OC-WL-TLlnC method using computer simulation results. In Fig. \ref{Fig7}, both the input noise $\boldsymbol{q}(k)$ and the output noise $p(k)$ are assumed to be zero-mean Gaussian with identical variances, $\sigma^2_q$ and $\sigma_p^2 = 0.1$. As shown in the figure, the simulated steady-state MSD follows the theoretical curve very closely, and the two results are almost indistinguishable at steady state. This strong agreement confirms that the derived steady-state MSD expression accurately captures the method behavior under Gaussian noise.
\FloatBarrier
\begin{figure}[!t]
    \centering
    \includegraphics[width=0.75\textwidth]{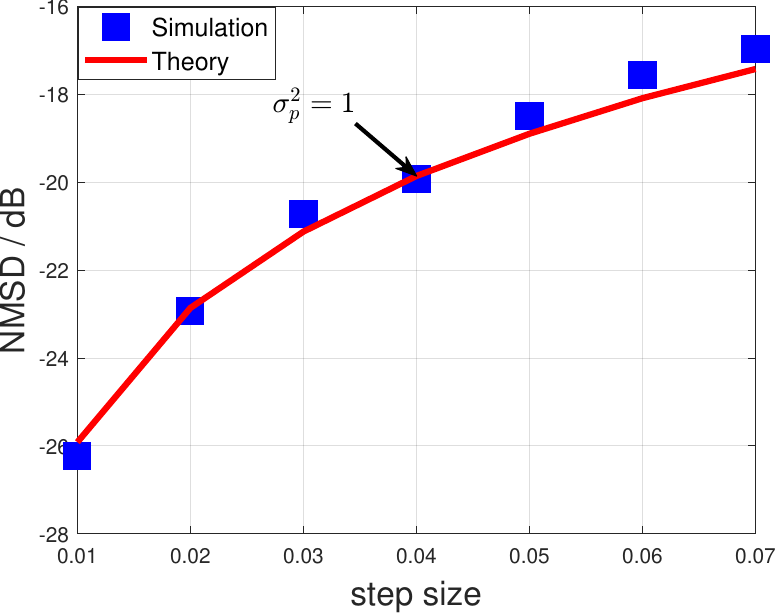}
    \caption{Comparison between theoretical and simulated steady-state MSD of the OC-WL-TLlnC algorithm under zero-mean Gaussian input noise ($\sigma_q^2 = 0.1$) and zero-mean Laplace output noise ($\sigma_p^2 = 1$).}
    \label{Fig8}
\end{figure}

To further assess the accuracy of the theoretical predictions, we also examine the steady-state MSD under generalized Gaussian noise in Fig. \ref{Fig8}. In this case, the input noise $\boldsymbol{q}(k)$ remains a zero-mean Gaussian with $\sigma_q^2 = 0.1$, whereas the output noise $p(k)$ follows a zero-mean Laplace distribution with variance $\sigma_p^2 = 1$, which is a special case of generalized Gaussian noise. As illustrated in Fig. \ref{Fig8}, the analytical steady-state prediction remains in close agreement with the simulation results, further validating the correctness of the steady-state analysis under non-Gaussian noise conditions.

    \section*{Credit authorship contribution statement}

    \textbf{Haiquan Zhao:} Writing– review editing, Investigation, Funding acquisition, Formal analysis, Data curation. \textbf{Kaleab Derbew Abebe:} Writing original draft, Validation, Supervision, Software, Methodology, Formal analysis, Data curation, Conceptualization. \textbf{Yi Peng:} Writing– review editing, Resource.

\section*{Declaration of competing interest}
None.
\section*{Acknowledgments}
This research received partial support from the National Natural Science Foundation of China under Grant 62171388, 61871461, 61571374.
\section*{Data Availability}
No data was used for the research described in the article.

\section{Conclusions}\label{sec:conclusions}
This work introduced the OC-WL-TLlnC method for power system frequency estimation. Building on the ACLlncosh framework, the proposed method addresses its limitations under persistent noise conditions by integrating WL modeling and OC into the TLlnC formulation. This approach enhances robustness by selectively discarding outlier data while preserving sensitivity to impulsive disturbances. Additional control parameters were incorporated to regulate the convergence speed and maintain adaptability in dynamic system environments. The simulation results under balanced and unbalanced conditions demonstrated that the OC-WL-TLlnC algorithm significantly outperforms the conventional ACLlncosh method, offering improved accuracy and stability in frequency tracking, especially during fault events and in the presence of noise.
\bibliographystyle{unsrt}
\bibliography{sections/mybibfile}
    \clearpage

    \clearpage

\end{document}